\def\Black{}
\def\Blue{}
\def\Brown{}
\documentstyle[aps,epsf,prl]{revtex}
\begin{document}
\begin{titlepage}
\null
\noindent
BARI-TH/00-372\\
CERN-TH/2000-30\\
DSF-2000/2\\
UGVA-DPT/2000-02-1070
\vspace{2cm}
\begin{center}
\Brown
\Large\bf 
Measuring $B\to\rho\pi$ decays and the unitarity angle $\alpha$
\Black
\end{center}
\vspace{1.5cm}

\begin{center}
\begin{large}
A. Deandrea$^a$, R. Gatto$^{b}$, M. Ladisa$^{a,c}$ G.
Nardulli$^{a,c}$ and P. Santorelli$^d$\\
\end{large}
\vspace{0.7cm}
$^a$ Theory Division, CERN, CH-1211 Gen\`eve 23, Switzerland\\
$^b$ D\'epartement de Physique Th\'eorique, Universit\'e de Gen\`eve,\\
24 quai E.-Ansermet, CH-1211 Gen\`eve 4, Switzerland\\
$^c$ Dipartimento di Fisica, Universit\`a di Bari and INFN Bari,\\
via Amendola 173, I-70126 Bari, Italy\\
$^d$ Dipartimento di Scienze Fisiche, Universit\`a di Napoli ``Federico II''\\
and INFN Napoli, Mostra d'Oltremare 20, I-80125 Napoli, Italy\\
\end{center}

\vspace{1.5cm}

\begin{center}
\begin{large}
\Brown
{\bf Abstract}\\[0.5cm]\Black
\end{large}
\parbox{14cm}{The decay mode $B\to\rho\pi$ is currently studied as a channel 
allowing, in principle, to measure without ambiguities the angle $\alpha$ of 
the unitarity triangle. It is also investigated by the
CLEO Collaboration where a branching ratio larger than expected for the
decay mode $B^\pm\to \rho^0\pi^\pm$ has been found. We investigate the role
that the $B^*$ and $B_0(0^+)$ resonances might play in these analyses.}
\end{center}

\vspace{2cm}
\noindent
\Blue
\Black

\noindent
January 2000
\end{titlepage}

\setcounter{page}{1}

%%%%%%%%%%%%%%%%%%%%%%%%%%%%%%%%%%%%%%%%%%%%%%
\preprint{CERN-TH/2000-30\\
BARI-TH/00-372\\
DSF-2000/2\\
UGVA-DPT/2000-02-1070}
\newcommand{\dd}{\displaystyle}
\newcommand{\nn}{\nonumber}
\newcommand{\be}{\begin{equation}}
\newcommand{\ee}{\end{equation}}
\newcommand{\bea}{\begin{eqnarray}}
\newcommand{\eea}{\end{eqnarray}}       
\title{Measuring $B\to\rho\pi$ decays and the unitarity angle $\alpha$}
\author{A. Deandrea$^a$, R. Gatto$^{b}$, M. Ladisa$^{a,c}$ G.
Nardulli$^{a,c}$ and P. Santorelli$^d$}
\address{$^a$ Theory Division, CERN, CH-1211 Gen\`eve 23, Switzerland}
\address{$^b$ D\'epartement de Physique Th\'eorique, Universit\'e de Gen\`eve,
24 quai E.-Ansermet, CH-1211 Gen\`eve 4, Switzerland}
\address{$^c$ Dipartimento di Fisica, Universit\`a di Bari and INFN Bari,
via Amendola 173, I-70126 Bari, Italy}
\address{$^d$ Dipartimento di Scienze Fisiche, Universit\`a di Napoli 
``Federico II'' and INFN Napoli,\\ 
Mostra d'Oltremare 20, I-80125 Napoli, Italy}
\date{January 2000}
\maketitle

\begin{abstract}
The decay mode $B\to\rho\pi$ is currently studied as a channel 
allowing, in principle, to measure without ambiguities the angle $\alpha$ of 
the unitarity triangle. It is also investigated by the
CLEO Collaboration where a branching ratio larger than expected for the
decay mode $B^\pm\to \rho^0\pi^\pm$ has been found. We investigate the role
that the $B^*$ and $B_0(0^+)$ resonances might play in these analyses.
\end{abstract}

\section{Introduction}    
The measurement of the angle $\alpha$ in the unitarity triangle will be
one of the paramount tasks of the future b--factories, such as the
dedicated $e^+e^-$ machines for the BaBar experiment at SLAC 
\cite{babar} and the BELLE experiment at KEK \cite{kek}, or 
hadron machines such as the LHC at CERN, with its program for
$B$--physics \footnote{Opportunities for $B$--physics at the
LHC have been recently discussed at the workshop on Standard Model
Physics (and More) at the LHC, 14-15 October 1999; copies of
transparencies can be found at the site
http://home.cern.ch/~mlm/lhc99/oct14ag.html.}. Differently from
the investigation of the $\beta$ angle, for which the $B \to J/\Psi K_s$ 
channel has been pinned up \cite{beta} and ambiguities can be 
resolved \cite{pavernew}, the task of
determining  the angle $\alpha$ is complicated by the problem of separating
two different weak hadronic matrix elements, each carrying its own weak phase.
The evaluation of these contributions, referred to in the literature as the 
{\it tree} ($T$) and the {\it penguin} ($P$) contributions, suffers from the
common theoretical uncertainties related to the estimate of composite
four-quark operators between hadronic states. For these estimates, 
only approximate schemes, such as the factorization approximation, exist 
at the moment, and for this reason several ingenuous schemes have been devised,
trying to disentangle $T$ and $P$ contributions. In general one tries to
exploit the fact that in the $P$ amplitudes only the isospin--$1/2$ 
\footnote{If one neglects electroweak penguins.} part of
the non--leptonic Hamiltonian is active \cite{gronau}; 
by a complex measurement involving several different isospin amplitudes,
one should be able to separate the two amplitudes and to get rid of the
ambiguities arising from the ill--known penguin matrix elements.

One of the favorite proposals involves the study of the reaction
$B \to \rho\pi$, i.e. six channels arising from the neutral $B$ decay:
\bea
&&\bar B^0~\to~\rho^+\pi^-~,\label{eq:2} \\
&&\bar B^0~\to~\rho^-\pi^+~,\label{eq:3} \\
&&\bar B^0~\to~\rho^0\pi^0~,\label{eq:4}
\eea
together with the three charge--conjugate channels, and the charged
decay modes:
\bea
&&B^-~\to~\rho^-\pi^0~,\label{eq:5} \\
&& B^-~\to~\rho^0\pi^-~,\label{eq:6}
\eea 
with two other charge--conjugate channels. Different strategies have been
proposed to extract the angle $\alpha$, either involving 
all the decay modes of a $B$ into a
$\rho\pi$ pair as well as three time--asymmetric quantities
measurable in the three channels for neutral $B$ decays 
\cite{lipkin,snyder,pham}, or attempting to measure only the
neutral $B$ decay modes by looking at the time-dependent asymmetries in
different regions of the Dalitz plot\footnote{In this way
the measurement of a decay mode with two neutral pions in the final 
state, eq. (\ref{eq:5}), can be avoided.}.

Preliminary to these analyses is the assumption that, using cuts in the
three invariant masses for the pion pairs, one can extract the $\rho$
contribution without significant background contaminations. The
$\rho$ has spin $1$, the $\pi$ spin $0$ as well as the initial $B$, and
therefore the $\rho$ has angular distribution $\cos ^2 \theta$
($\theta$ is the angle of one of the $\rho$ decay products with the
other $\pi$ in the $\rho$ rest frame). This means that the Dalitz plot
is mainly populated at the border, especially the corners, by this
decay. Only very few events should be lost by excluding the interior of the
Dalitz plot, which is considered a good way to exclude or at least
reduce backgrounds. Analyses following these hypotheses were
performed by the BaBar working groups \cite{babar}; MonteCarlo
simulations, including the background from the $f_0$ resonance, show
that, with cuts at $m_{\pi\pi}=m_\rho\pm 300$ MeV, no significant contributions
from other sources are obtained. Also the role of excited resonances such as
the $\rho^\prime$ and the non--resonant background has been discussed
\cite{charles}.

A signal of possible difficulties for this strategy arises from new results
from the CLEO Collaboration recently reported at the DPF99 and APS99
Conferences \cite{gao}: 
\begin{equation} {\mathcal B}( B^\pm~\to~\rho^0\pi^\pm)~=~(1.5\pm 0.5\pm
0.4)\times 10^{-5}~,\label{eq:12}
\end{equation}
\begin{equation}
{\mathcal B}( B~\to~\rho^\mp\pi^\pm)~=~(3.5^{+1.1}_{-1.0}\pm
0.5)\times 10^{-5}~,\label{eq:13}
\end{equation} 
with a ratio
\begin{equation}\label{eq:88}
R~=~\frac{ {\mathcal B}( B~\to~\rho^\mp\pi^\pm) }{ {\mathcal B}(
B^\pm~\to~\rho^0\pi^\pm)}~=~2.3\pm 1.3~.
\end{equation}
As discussed in \cite{gao}, this ratio looks rather small; as a matter of
fact, when computed in simple approximation schemes, including
factorization with no penguins, one gets, from the  DDGN model of
Ref. \cite{deandrea}, $R\simeq 13$, 
admittedly with a large uncertainty; another popular
approach, i.e. the WBS model \cite{WBS}, gives $R \simeq 6$ (in both 
cases we use $a_1=1.02$, $a_2=0.14$).
The aim of the present study is to show that a new contribution,
not discussed before, is indeed relevant to the decay
(\ref{eq:6}) and to a lesser extent to the decay (\ref{eq:4}). It
arises from the virtual resonant production
depicted in Fig. 1, where the intermediate particle is the $B^*$
meson resonance or other excited states. The $B^*$ resonance, 
because of phase--space limitations, cannot be produced on the mass shell.
Nonetheless the $B^*$ contribution might be important, owing to its almost
degeneracy in  mass with the $B$ meson; therefore its  tail 
may produce sizeable effects in some of the decays of $B$
into light particles, also because it is known theoretically that the
strong coupling constant between $B$, $B^*$ and a pion is
large \cite{gatto}. Concerning other states, we expect their role to
decrease with their mass, since there is no enhancement from the virtual
particle propagator; we shall only consider the $0^+$ state $B_0$ with
$J^P=0^+$ because its coupling to a pion and the meson $B$ is known
theoretically to be uniformly (in momenta) large \cite{gatto}.
The plan of the paper is as follows. In Section \ref{II} we list the
hadronic quantities that are needed for the computation of the widths; in
Section \ref{III} we present the results and finally, in Section
\ref{IV}, we give our conclusions.

\section{Matrix elements\label{II}}

The effective weak non-leptonic Hamiltonian for the $|\Delta
B|=1$ transition is \footnote{We omit, as 
usual in these analyses, the electroweak operators
$Q_k$ ($k=$ 7, 8, 9, 10); they are in general small, but for 
$Q_9$, whose role might be sizeable; its inclusion in
the present calculations would be straightforward.}:
\be
H~=~\frac{G_F}{\sqrt 2}\left\{ V^*_{ub}V_{ud}\sum_{k=1}^2
C_k(\mu)Q_k~-~V^*_{tb}V_{td}\sum_{k=3}^6
C_k(\mu)Q_k\right\}~.\ee
The operators relevant to the present analysis
are the so--called current--current operators:
\bea
Q_1&=&
(\bar d_\alpha u_\beta )_{V-A}
(\bar u_\beta b_\alpha )_{V-A}\nn\\
Q_2&=&
(\bar d_\alpha u_\alpha )_{V-A}
(\bar u_\beta b_\beta )_{V-A}~,
\eea
and the QCD penguin operators:
\bea
Q_3&=&
(\bar d_\alpha b_\alpha )_{V-A}\sum_{q^\prime=u,d,s,c,b}
(\bar q^\prime_\beta q^\prime_\beta  )_{V-A}\nn\\
Q_4&=&
(\bar d_\alpha b_\beta )_{V-A}\sum_{q^\prime=u,d,s,c,b}
(\bar q^\prime_\beta q^\prime_\alpha  )_{V-A}\nn\\
Q_5&=&
(\bar d_\alpha b_\alpha )_{V-A}\sum_{q^\prime=u,d,s,c,b}
(\bar q^\prime_\beta q^\prime_\beta  )_{V+A}\nn\\
Q_6&=&
(\bar d_\alpha b_\beta )_{V-A}\sum_{q^\prime=u,d,s,c,b}
(\bar q^\prime_\beta q^\prime_\alpha  )_{V+A}~,
\eea
We use the following values of the Wilson coefficients:
$C_1=-0.226$, $C_2=1.100$, $C_3=0.012$, $C_4=-0.029$, $C_5=0.009$,
$C_6=-0.033$; they are obtained in the HV scheme \cite{buras}, with
$\Lambda^{(5)}_{\bar{MS}}=225$ MeV, $\mu={\bar m}_b(m_b)=4.40$ GeV and
$m_t=170 $ GeV.
For the CKM mixing matrix \cite{ckm} we use the Wolfenstein parameterization
\cite{wfstn} with $\rho=0.05$, $\eta = 0.36$ and $A=0.806$ in the 
approximation accurate to order $\lambda^3$ in the real part and
$\lambda^5$ in the imaginary part, i.e. $V_{ud}=1-\lambda^2 /2$,
$V_{ub}=A \lambda^3 \;[\rho - i \eta \; (1-\lambda^2 /2)]$, $V_{td}=
A \lambda^3 (1-\rho - i \eta)$ and $V_{tb}=1$.

The diagram of Fig.~1 describes two processes.
For the $B^*$ intermediate state there is an emission of a  
pion by strong interactions, followed by the weak decay of the virtual
$B^*$ into two pions; for the $\rho$ intermediate state there is a
weak decay of $B \to \rho \pi$ followed by the strong decay of 
the $\rho$ resonance. We compute these diagrams as Feynman graphs
of an effective theory within the factorization approximation, 
using information from the effective Lagrangian
for heavy and light mesons and form factors for the couplings to the
weak currents \footnote{In the second reference of \cite{pavernew} a 
similar approach has been used to describe the decay mode 
$B^0\to D^+D^-\pi^0$; the main difference is that for $B \to 3 \pi$ we 
cannot use soft pion theorems and chiral perturbation
theory, because the pions are in general hard; therefore we have to use
information embodied in the semileptonic beauty meson form factors. This
is also the main difference with respect to \cite{pham}.}.

To start with we consider the strong
coupling constants. They are defined as 
\bea
\langle {\bar B}^0(p^\prime)~\pi^-(q)~|~B^{*\, -}(p,\epsilon)\rangle
~&=&~g^{B^* B\pi}\epsilon\cdot q\nn\\
\langle B^-(p^\prime)~\pi^+ (q)~|~{\bar B}^0_0(p)\rangle~
&=&~G^{B_0 B\pi}(p^2)\\
\langle \pi^0(q^\prime)~\pi^-(q)~|~\rho^-(p,\epsilon)\rangle 
~&=&~g_\rho \epsilon\cdot (q^\prime - q)\nn .
\eea
In the heavy quark mass limit one has
\bea
g^{B^* B\pi}~&=&~\frac{2 m_B g}{f_\pi}\\
G^{B_0 B\pi}(s)~&=&~-~
\sqrt{\frac{m_{B_0} m_B }{2}}\frac{s-m^2_B}{m_{B_0}}
\frac{h}{f_\pi}~.
\eea
For $g$ and $h$ we have limited experimental information and we have
to use some theoretical inputs. For $g$ and $h$  reasonable ranges of
values are $g=0.3$--$0.6$, $-h=0.4$--$0.7$ \cite{casalbuoni}. 
These numerical estimates encompass results 
obtained by different methods:
QCD sum rules \cite{gatto},  potential models \cite{defazio}, effective
Lagrangian \cite{casalbuoni0}, NJL-inspired models \cite{polosa0}.
Moreover $g_\rho=5.8$ and $f_\pi\sim 130$ MeV. This value of $g_\rho$ 
is commonly used in the chiral effective
theories including the light vector meson resonances 
and corresponds to $\Gamma_\rho\simeq 150$ MeV; see, for instance, 
\cite{casalbuoni}, where a review of different methods for the
determination of $g$ is also given.

For the matrix elements of quark bilinears between hadronic states, we
use the following matrix elements:
\bea
\langle \pi^-~|{\bar d}\gamma_5 u|~0\rangle
&=&\frac{i f_\pi m_\pi^2 }{2m_q}\nn\\
\langle \pi^0(q)~|{\bar u}\gamma_5 b|~{B}^{*\, -}(p)\rangle 
&=& i \epsilon^\mu (q-p)_\mu
\frac{2 m_{B^*}A^\pi_0 }{m_b+ m_q}\nn\\
\langle \rho^+(q,~\epsilon)~|{\bar u}\gamma_5 b|~{\bar
B}^{0}(p)\rangle&=& i \epsilon^{*\, \mu} (p-q)_\mu
\frac{2 m_{\rho}A_0 }{m_b+ m_q}\nn\\
\langle \pi^+(q)~|{\bar u}\gamma_\mu b|~{\bar B}^{0}(p) \rangle &=&
F_1\left[(p+q)^\mu-\frac{m^2_B-m^2_\pi}{(p-q)^2}(p-q)^\mu\right]
+F_0\, \frac{m^2_B-m^2_\pi}{(p-q)^2}(p-q)^\mu\nn\\
\langle \pi^+(q)~|{\bar u}\gamma^\mu(1-\gamma_5) b|~{\bar B}_0^{0}(p)\rangle 
&=& i \left\{ {\tilde
F}_1\left[(p+q)^\mu-\frac{m^2_{B_0}-m^2_\pi}{(p-q)^2}(p-q)^\mu\right]
+{\tilde F}_0\, \frac{m^2_{B_0}-m^2_\pi}{(p-q)^2}(p-q)^\mu \right\} \nn\\
\langle 0~|{\bar u}\gamma^\mu d|~\rho^-(q,~\epsilon)\rangle 
&=&f_\rho \epsilon^{ \mu}~,
\eea 
where $f_\rho=0.15 {\rm ~ GeV}^2$ \cite{bft} and 
\bea
A_0^\pi&=&A_0^\pi(0)~=~0.16 \; , \; \; \; \; \; \; \; \;\hskip 1.0cm
A_0=A_0(0)=0.29~,\label{eq:52}\\
F_1&=&F_1(0)=F_0(0)=0.37 \; , \; \; \; \; \; \; \; \;
F_0^{\pi} = {\tilde F}_1(0)={\tilde F}_0(0)=-0.19\label{eq:54}~.
\eea 
The first three numerical inputs have been obtained by the relativistic
potential model; $A_0$ and $F_1$ can be found in
\cite{santorelli}, while $A_0^\pi$ has been obtained here for the first
time, using the same methods. The last figure in (\ref{eq:54}) concerns 
$F_0^{\pi}$, for which such an 
information is not available; for it  we used the methods 
of \cite{casalbuoni} and the strong
coupling $B B_0 \pi$ computed in \cite{polosa}.

\section{Amplitudes and numerical results\label{III}}

For all the channels we consider three different contributions
$A_\rho$, $A_{B^*}$, $A_{B_0}$, due respectively to the $\rho$ 
resonance, the $B^*$ pole and the $B_0$
positive parity $0^+$ resonance, whose mass we take \footnote{ 
We identify the $0^+$ state mass with the average mass of the $B^{**}$
states given in \cite{pdg}.} to be $5697$ MeV. 

For each of the amplitudes
\bea
A^{--+}~&=&~A(B^-\to\pi^-\pi^-\pi^+)~\label{eq:41}\\
A^{-00}~&=&~A(B^-\to\pi^-\pi^0\pi^0)~\label{eq:42}\\
A^{+-0}~&=&~A({\bar B}^0\to\pi^+\pi^-\pi^0)~\label{eq:43}
\eea
we write the general formula\footnote{We add coherently the three
contributions; the relative sign the $B$ resonances on one side and the
$\rho$ contribution on the other is irrelevant, as the former are
dominantly real and the latter is dominantly imaginary. The relative sign
between $B^*$ and $B_0$ is fixed by the effective Lagrangian for heavy
mesons.} $A^{ijk} = A^{ijk}_\rho
+ A^{ijk}_{B^*}+ A^{ijk}_{B_0}$. We get, for the process (\ref{eq:41}):
\bea
A^{--+}_\rho&=&\, {\bar\eta}^0\,\left[ \frac{
t^\prime-u}{t-m^2_\rho+i\Gamma_\rho
m_\rho}~+ \frac{
t-u}{t^\prime-m^2_\rho+i\Gamma_\rho
m_\rho} \right]~\nn \\
A^{--+}_{B^*}&=& K\, \left[
\frac{\Pi(t,u)}{t-m^2_{B^*}+i\Gamma_{B^*}m_{B^*}}
~+~ \frac{\Pi(t^\prime,u) }{ t^\prime
-m^2_{B^*}+i\Gamma_{B^*}m_{B^*} } \right]~, \nn\\
A^{--+}_{B0}&=& {\tilde K}^0 \, (m_{B_0}^2 - m_\pi^2 ) \, 
\left[\frac{1}{t-m^2_{B_0}+i\Gamma_{B_0} m_{B_0}} +
\frac{1}{t^\prime -m^2_{B_0}+i\Gamma_{B_0} m_{B_0}} \right]~,
\eea
where, if $p_{\pi^-}$ is the momentum of one of the two negatively
charged pions $t=(p_{\pi^-}+p_{\pi^+})^2$, 
$t^\prime$ is obtained by exchanging the two identical pions and
$u$ is the invariant mass of the two identical negatively charged
pions. Clearly one has $u+t+t^\prime~=~m_B^2+3m_\pi^2$. The expressions
entering in the previous formulas are
\bea
{\bar \eta}^0&=&\, \frac{G_F}{\sqrt 2} V_{ub}V^*_{ud} 
\frac{g_\rho }{\sqrt 2}
\left[ f_\rho F_1 \left(c_1+\frac{c_2}{3}\right)~+m_\rho A_0
f_\pi\left(c_2+\frac{c_1}{3}\right)\right] ~+ \nn\\
&+& \frac{G_F}{\sqrt 2} V_{tb}V^*_{td}
\frac{g_\rho }{\sqrt 2}\left[ \left(c_4+\frac{c_3}{3}\right)
\left(f_\rho F_1\, -\, m_\rho A_0
f_\pi\right)~+~2\, \left(c_6+\frac{c_5}{3}
\right) m_\rho A_0
f_\pi\frac{m_\pi^2}{(m_b+m_q)2 m_q}\right]~,\nn \\
K&=&- 4 {\sqrt 2} \; g \; m^2_B A_0^\pi \frac{G_F}{\sqrt 2} \, \Big\{
V_{ub}V^*_{ud}\left(c_2+\frac{c_1}{3}\right) -
V_{tb}V^*_{td}\left[c_4+\frac{c_3}{3}-2\left(c_6+\frac{c_5}{3}\right)
\frac{m_\pi^2}{(m_b+m_q)2 m_q}\right]
\Big\} \nn \\
{\tilde {K}}^0 &=& h {\sqrt{\frac{m_B}{m_{B0}}}} F_0^\pi 
{\frac{G_F}{\sqrt 2}} (m^2_{B0}-m^2_B) \,
\Big\{ V_{ub}V^*_{ud}\left(c_2+\frac{c_1}{3}\right)
- V_{tb}V^*_{td}\left[c_4+\frac{c_3}{3}-2\left(c_6+\frac{c_5}{3}\right)
\frac{m_\pi^2}{(m_b+m_q)2 m_q}\right] \Big\} ~,
\label{ktildezero}
\eea with $m_b= 4.6$ GeV, $m_q\sim m_u\sim m_d\simeq 6$ MeV,
$\Gamma_{B^*}=0.2$ keV,  $\Gamma_{B_0}=0.36$ GeV \cite{casalbuoni}.
Moreover, for the process (\ref{eq:42})
\bea
A^{-00}_\rho&=&\, {\bar\eta}^-\,\left[ \frac{
s^\prime-u}{s-m_{\rho}^2+i\Gamma_\rho \; m_\rho}~+ \frac{
s-u}{s^\prime-m^2_\rho+i\Gamma_\rho \; m_\rho} \right]~,
\nn\\
A^{-00}_{B^*}&=&\, \frac{1}{\sqrt{2}}\, \Big\{
\, K_1\frac{s+s^\prime-4m_\pi^2}{2m^2_{B^*}}
+\frac{K\, \Pi(s^\prime,s)~+~K_1\,
\Pi(s^\prime,u) }{s^\prime-m^2_{B^*}+i\Gamma_{B^*}m_{B^*}}~+~
\frac{K\, \Pi(s,s^\prime)~+~K_1\,
\Pi(s,u)}{s-m^2_{B^*}+i\Gamma_{B^*}m_{B^*}} \Big\}~, \nn \\
A^{-00}_{B_0}&=& \, 
\left(
\frac{{\tilde K}^{0} +{\tilde K}^{cc}}{s-m^2_{B_0}+i\Gamma_{B_0} m_{B_0}}
+ 
\frac{{\tilde K}^{0} +{\tilde K}^{cc}}{s^\prime
-m^2_{B_0}+i\Gamma_{B_0} m_{B_0}}
+ \frac{{\tilde K}^{0}}{u-m^2_{B_0}+i\Gamma_{B_0} m_{B_0}}\right)\; 
\frac{(m_{B_0}^2 -m_\pi^2 )}{2}
\eea 
In this case we define $s=(p_{\pi^-}+p_{\pi^0})^2$,
if $p_{\pi^0}$ is the momentum of one of the  two identical neutral pions,
$s^\prime$ is obtained by exchanging the two neutral pions and
$u$ is their invariant mass (again we have a relation among the different
Mandelstam variables: $s+s^\prime+u=m^2_B+3 m^2_\pi$).
Then ${\bar\eta}^-$, $K_1$ and ${\tilde {K}}^{cc}$ are given by
\bea
{\bar \eta}^-&=&\, \frac{G_F}{\sqrt 2} V_{ub}V^*_{ud}
\frac{g_\rho }{\sqrt 2}
\left[ f_\rho F_1 \left(c_2+\frac{c_1}{3}\right) + m_\rho A_0
f_\pi\left(c_1+\frac{c_2}{3}\right)\right]~+ \nn\\
&+& \frac{G_F}{\sqrt 2} V_{tb}V^*_{td}
\frac{g_\rho }{\sqrt 2}\left[ \left(c_4+\frac{c_3}{3}\right)
\left(-f_\rho F_1\, +\, m_\rho A_0
f_\pi\right)~-~2\, \left(c_6+\frac{c_5}{3}
\right) m_\rho A_0
f_\pi\frac{m_\pi^2}{(m_b+m_q)2 m_q}\right]~,\nn\\
K_1&=&  - 4 \; g \; m^2_B A_0^\pi\, \frac{G_F}{\sqrt 2}\Big\{
V_{ub}V^*_{ud}\left(c_1+\frac{c_2}{3}\right) +
V_{tb} V^*_{td}\left[ c_4+\frac{c_3}{3}-2\left(c_6+\frac{c_5}{3}\right)
\frac{m_\pi^2}{(m_b+m_q)2 m_q}\right]
\Big\}~,\nn \\
{\tilde {K}}^{cc} &=& h {\sqrt{\frac{m_B}{m_{B0}}}} F_0^\pi 
{\frac{G_F}{\sqrt {2}}} (m^2_{B0}-m^2_B) \,
\Big\{ V_{ub}V^*_{ud}\left(c_1+\frac{c_2}{3}\right)
+ V_{tb}V^*_{td}\left[c_4+\frac{c_3}{3}-2\left(c_6+\frac{c_5}{3}\right)
\frac{m_\pi^2}{(m_b+m_q)2 m_q}\right]
\Big\}~,
\eea
and
\be
\Pi(x,y)= m_\pi^2-\frac{y}{2}+\frac{x(m_B^2-m^2_\pi-x)}{4m_{B^*}^2}~,
\ee
while ${\tilde {K}}^0$ was given above in (\ref{ktildezero}).

Finally, for the neutral $B$ decay (\ref{eq:43}), we have
\bea
A^{+-0}_\rho&=&\eta^0\, \frac{
u-s}{t-m^2_\rho+i\Gamma_\rho \; m_\rho}~+~\eta^+\,\frac{
s-t}{u-m^2_\rho+i\Gamma_\rho \; m_\rho}~+~\eta^-\, \frac{
t-u}{s-m^2_\rho+i\Gamma_\rho \; m_\rho} \nn\\
A^{+-0}_{B^*}&=& ~\frac{K\, \Pi(s,t)~+~K_1\,
\Pi(s,u) }{s - m^2_{B^*} + i \Gamma_{B^*} m_{B^*} }~
-~\frac{ K\, \Pi(t,s) }{ t- m^2_{B^*} + i \Gamma_{B^*} m_{B^*}} \nn\\
A^{+-0}_{B_0}&=& \, \left(\frac{{\tilde K}^{0} +{\tilde K}^{cc}}
{s-m^2_{B_0}+i\Gamma_{B_0} m_{B_0}} + \frac{{\tilde K}^{0}}
{t-m^2_{B_0}+i\Gamma_{B_0} m_{B_0}} \right) \; 
(m_{B_0}^2 -m_\pi^2 )~,
\eea
where $s=(p_{\pi^-}+p_{\pi^0})^2$, $t=(p_{\pi^-}+p_{\pi^+})^2$,
$u=(p_{\pi^+}+p_{\pi^0})^2$, and $s+t+u = m_B^2+3m_\pi^2$.
The constants appearing in these equations are :
\bea
\eta^0&=&- \frac{g_\rho}{2}( f_\rho F_1+m_\rho A_0
f_\pi)\frac{G_F}{\sqrt 2}\left[
V_{ub}V^*_{ud}\left(c_1+\frac{c_2}{3}\right) +
V_{tb}V^*_{td}\left(c_4+\frac{c_3}{3}\right)\right]~+\nn\\
&+&  g_\rho m_\rho A_0
f_\pi  \,  \frac{G_F}{\sqrt 2}\,
V_{tb}V^*_{td}\left(c_6+\frac{c_5}{3}
\right)\frac{m_\pi^2}{(m_b+m_q)2 m_q} \nn\\
\eta^+&=& g_\rho m_\rho A_0
f_\pi\frac{G_F}{\sqrt 2}\left\{
V_{ub}V^*_{ud}\left(c_2+\frac{c_1}{3}\right) ~-~
V_{tb}V^*_{td}\left[c_4+\frac{c_3}{3}-2\left(c_6+\frac{c_5}{3}\right)
\frac{m_\pi^2}{(m_b+m_q)2 m_q}\right]
\right\} \nn\\
\eta^-&=& g_\rho f_\rho F_1\frac{G_F}{\sqrt 2}
\left\{V_{ub}V^*_{ud}\left(c_2+\frac{c_1}{3}\right) ~-~
V_{tb}V^*_{td}\left(c_4+\frac{c_3}{3} \right) \right\}~.
\eea

For the charged $B$ decays we obtain the results in  Table I and II. In
order to show the dependence of the results on the numerical
values of the different input parameters, we consider in Table I results
obtained with $g=0.40$ and $h=-0.54$, which lie in the middle of the
allowed ranges, while in Table II we present the results obtained with  
$g=0.60$ and $h=-0.70$, which represent in a sense an extreme case (we do
not consider the dependence on other numerical inputs, e.g. form factors,
which can introduce further theoretical uncertainty). In both cases  the
branching ratios are obtained with $\tau_B=1.6$ psec and, by integration
over a limited section of the Dalitz plot, defined as $m_\rho-\delta\leq
({\sqrt t}, {\sqrt t^\prime}) \leq m_\rho+\delta$ for $B^-\to\pi^-\pi^-\pi^+$  
and $m_\rho-\delta\leq ({\sqrt s}, {\sqrt s^\prime} ) \leq m_\rho+\delta$ 
for $B^-\to\pi^-\pi^0\pi^0$. For $\delta$ we take $300$ MeV. This amounts 
to require that two of the three pions (those corresponding to the charge 
of the $\rho$) reconstruct the $\rho$ mass within an interval of $2 \delta$. 
Numerical uncertainty due to the integration procedure is $\pm 5\%$.

We can notice that the inclusion of the new diagrams 
($B$ resonances in Fig. 1) produces practically no effect
for the $B^-\to\pi^-\pi^0\pi^0~$  decay mode, while for 
$B^-\to\pi^+\pi^-\pi^-~$ the effect is significant. 
For the choice of parameters in Table I the overall effect is an increase 
of $50\%$ of the branching ratio as compared to the result
obtained by the $\rho$ resonance alone. In the case of Table II we
obtain an even larger result, i.e.  a total branching ratio ${\mathcal
B}(B^-\to\pi^+\pi^-\pi^-)$ of $0.82\times 10^{-5}$, in reasonable 
agreement with the experimental result 
(\ref{eq:12}) (the contribution of the $\rho$ alone would produce a
result smaller by a factor of 2). It should be observed that the events
arising from the $B$ resonances diagrams represent an irreducible
background, as one can see from the sample Dalitz plot depicted in Fig. 2
for the $B^-\to\pi^+\pi^-\pi^-~$ (on the axis the two $m^2_{\pi^+\pi^-}$
squared invariant masses). The
contributions from the $B$ resonances populate the whole Dalitz plot and,
therefore, cutting around $t\sim t^\prime\sim m_\rho$ significantly
reduces them. Nevertheless their effect can survive the experimental cuts,
since there will be enough data at the corners, where the contribution 
from the $\rho$ dominates. Integrating on the whole Dalitz plot, with no cuts
and including all contributions, gives:
\begin{equation}
{\mathrm Br} (B^-\to\pi^-\pi^0\pi^0) = 1.5 \times 10^{-5} \;\;\;\;\;\;
{\mathrm Br} (B^-\to\pi^+\pi^-\pi^-) = 1.4 \times 10^{-5}
\end{equation}
where the values of the coupling constants are as in Table I.

We now turn to the neutral $B$ decay modes. We define effective width
integrating the Dalitz plot only in a region around the $\rho$ resonance:
\begin{eqnarray}
\Gamma_{eff}({\bar B}^0\to\rho^-\pi^+)&=&
\Gamma({\bar B}^0\to\pi^+\pi^-\pi^0)\Big|_{ m_\rho-\delta
\leq {\sqrt s}\leq m_\rho+\delta }\\
\Gamma_{eff}({\bar B}^0\to\rho^+\pi^-)&=&
\Gamma({\bar B}^0\to\pi^+\pi^-\pi^0)\Big|_{ m_\rho-\delta\leq
{\sqrt u}\leq m_\rho+\delta }\\
\Gamma_{eff}({\bar B}^0\to\rho^0\pi^0)&=&
\Gamma({\bar B}^0\to\pi^+\pi^-\pi^0)\Big|_{m_\rho-\delta\leq
{\sqrt t}\leq m_\rho+\delta}~.
\end{eqnarray}
The Mandelstam variables have been defined above and again we use
$\delta=300$ MeV \cite{libby}. Similar definitions hold for the $B^0$ 
decay modes. The results in Table III show basically no
effect for the $\bar{B}^0\to\rho^\pm\pi^\mp$ decay channels and a moderate
effect for the $\rho^0\pi^0$ decay channel. The effect in this channel 
is of the order of 20\% (resp. 50\%) for ${\bar B}^0$ (resp. $B^0$) decay,  
for the choice $g=0.60, \, h=-0.70$; for smaller
values of the strong coupling constants the effect is reduced.
Integration on the whole Dalitz plot, including all contributions, gives 
\begin{equation}
{\mathrm Br} (\bar{B}^0\to \pi^+\pi^-\pi^0) = 2.6 \times 10^{-5}
\end{equation}
confirming again that most of the branching ratio is due to the
$\rho$-exchange (the first three lines of the $\rho$
column in Table III sum up to $2.3 \times 10^{-5}$).
 
To allow the measurement of $\alpha$,
the experimental programmes will consider
the asymmetries arising from the time--dependent amplitude:
\begin{equation}
{\mathcal A} (t)=e^{-\Gamma/2 t}\left(\cos \frac{\Delta m t}{2} 
A^{+-0}\pm i \sin \frac{\Delta mt}{2} {\bar A}^{+-0} \right)~,
\end{equation}
where one chooses the $\pm$ sign according to the flavor of
the $B$, and $\Delta m $ is the mass difference between the two mass
eigenstates in the neutral $B$ system. Here ${\bar A}^{+-0}$ is the 
charge--conjugate amplitude. We have performed asymmetric
integrations over the Dalitz plot for three variables: $R_1$, $R_2$ and
$R_3$, which multiply, in the time--dependent asymmetry, respectively $1$, 
$\cos\Delta m t$ and $\sin\Delta m t$. We have found no significant effect 
due to the $B^*$ or the $B_0$ resonance for $R_1$ and $R_3$. On the other
hand these effects are present in $R_2$, but $R_2$ is
likely to be too small to be accurately measurable. 

\section{Conclusions\label{IV}}    

In conclusion our analysis shows that the effect of including $B$
resonance polar diagrams is significant for the $B^-\to\pi^-\pi^-\pi^+$
and negligible for the other charged $B$ decay mode. This result is of
some help in explaining the recent results from the CLEO Collaboration,
since we obtain
\begin{equation}
R~=~3.5\pm 0.8~,
\end{equation}
to be compared with the experimental result in eq. (\ref{eq:88}). The $\rho$
resonance alone would produce a result  up to a factor of 2 higher. 
Therefore we conclude that the polar diagrams examined in this paper 
are certainly relevant in the study of the charged $B$ decay into three pions.

In the case of neutral $B$ decays we have found that, as far as the
branching ratios are concerned, the only decay mode where the contribution
from the fake $\rho$'s (production of a pion and the $B^*$ or
the $B_0$ resonance) may be significant is the neutral $\rho^0\pi^0$ decay
channel. As for the time--dependent asymmetry no significant
effect is found. Therefore the $B\to\pi\pi\pi$ decay channel 
allows an unambiguous measurement of
$\alpha$, with two provisos: 1) only the neutral $B$ decay modes
are considered; 2) the $\rho^0\pi^0$ final state can be disregarded from
the analysis.

\acknowledgments{We thank J. Charles, Y. Gao, J. Libby, A. D. Polosa
and S. Stone for discussions.}

\twocolumn

\begin{figure}
\epsfxsize=7cm
\centerline{\epsffile{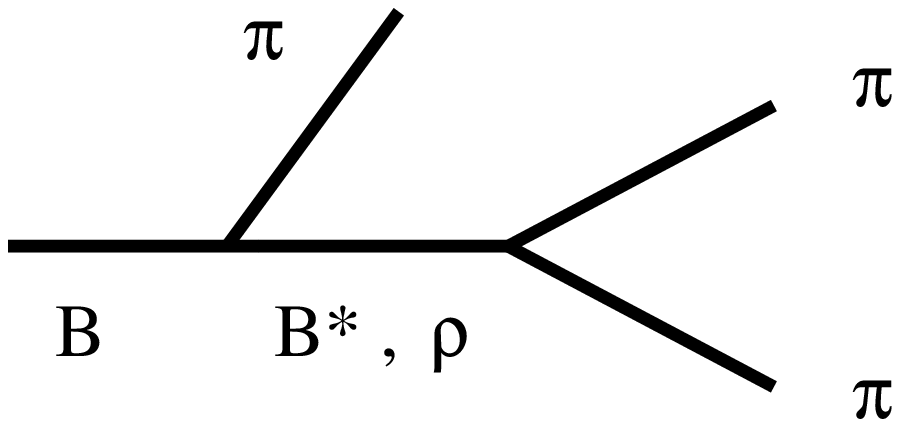}}
\noindent
{\bf Fig. 1} - {The polar diagram. For the $B$ resonances ($B^*=1^-$, $0^+$) 
the strong coupling is on the left and the weak coupling on the right; the 
situation is reversed for the $\rho$ production.}
\end{figure}

\begin{figure}
\epsfxsize=7cm
\centerline{\epsffile{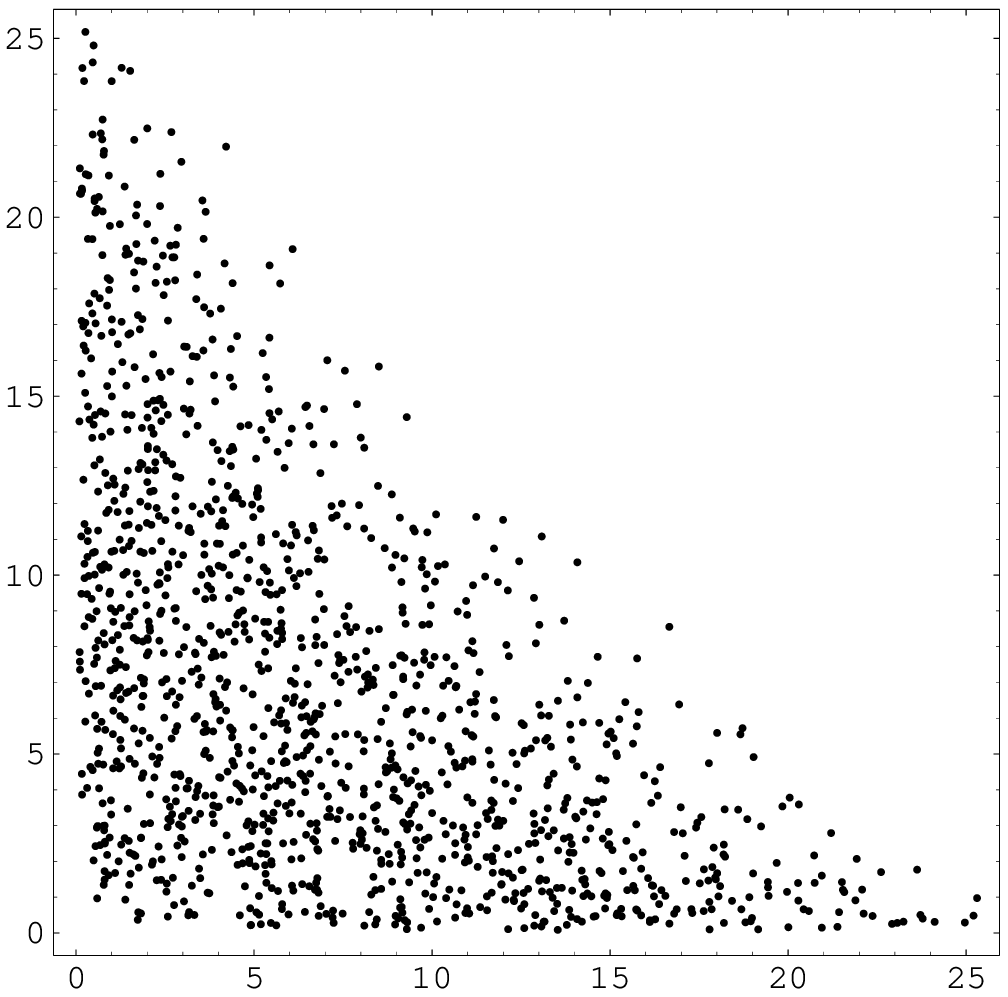}}
\noindent
{\bf Fig. 2} - {Sample Dalitz plot for the decay $B^- \to \pi^+ \pi^-
\pi^-$. In order to show the mass distribution of the $B$ resonance
diagrams, only their
contribution is taken into account
for this plot.}
\end{figure}

\narrowtext
\begin{table}
\caption{Effective branching ratios for the charged $B$ decay channels into 
three pions for the choice of the strong coupling constants $g=0.40$ and 
$h=-0.54$. Cuts as indicated in the text.} 
\begin{tabular}{cccc}
Channels& $\rho$ & $\rho+B^*$ & $\rho+B^*+ B_0$ \\
$B^-\to\pi^-\pi^0\pi^0~$ &
$1.0\times 10^{-5}$&
$1.0\times 10^{-5}$&
$1.0\times 10^{-5}$ \\
$B^-\to\pi^+\pi^-\pi^-$& 
$0.41\times 10^{-5}$&
$0.58\times 10^{-5}$&
$0.63\times 10^{-5}$
\end{tabular}
\end{table}

\begin{table}
\caption{Effective branching ratios for the charged $B$ decay channels into 
three pions for the choice of the strong coupling constants $g=0.60$ and 
$h=-0.70$. Cuts as indicated in the text.} 
\begin{tabular}{cccc}
Channels& $\rho$ & $\rho+B^*$ & $\rho+B^*+ B_0$ \\
$B^-\to\pi^-\pi^0\pi^0~$ &
$1.1\times 10^{-5}$&
$1.0\times 10^{-5}$&
$1.1\times 10^{-5}$ \\
$B^-\to\pi^+\pi^-\pi^-$& 
$0.41\times 10^{-5}$&
$0.74\times 10^{-5}$&
$0.82\times 10^{-5}$
\end{tabular}
\end{table}
\begin{table}
\caption{Effective branching ratios for the neutral $ B$ decay
channels into $\rho\pi$ ($g=0.60,\, h=-0.70$). Cuts as indicated in the text.} 
\begin{tabular}{cccc}
Channels& $\rho$ & $\rho+B^*$ & $\rho+B^*+ B_0$ \\
${\bar B}^0\to\rho^-\pi^+~$ &
$0.50\times 10^{-5}$&
$0.52\times 10^{-5}$&
$0.49\times 10^{-5}$ \\
${\bar B}^0\to\rho^+\pi^-~$ &
$1.7\times 10^{-5}$&
$1.7\times 10^{-5}$&
$1.7\times 10^{-5}$ \\
${\bar B}^0\to\rho^0\pi^0~$ &
$0.10\times 10^{-5}$&
$0.15\times 10^{-5}$&
$0.12\times 10^{-5}$ \\ \hline
${B}^0\to\rho^+\pi^-~$ &
$0.49\times 10^{-5}$&
$0.51\times 10^{-5}$&
$0.48\times 10^{-5}$ \\
${B}^0\to\rho^-\pi^+~$ &
$1.7\times 10^{-5}$&
$1.7\times 10^{-5}$&
$1.7\times 10^{-5}$ \\
${B}^0\to\rho^0\pi^0~$ &
$0.11\times 10^{-5}$&
$0.17\times 10^{-5}$&
$0.15\times 10^{-5}$ 
\end{tabular}
\end{table}
\end{document}